\documentclass[aps, prd,twocolumn,nofootinbib,tightenlines,floatfix,showpacs ,amssymb,superscriptaddress]{revtex4-2}
\usepackage{amsmath,graphicx}
\usepackage{amssymb}
\usepackage{mathrsfs}
\usepackage{lipsum}
\usepackage{indentfirst}
\usepackage{cases}
\usepackage{graphicx}
\usepackage{subfig}
\usepackage{booktabs}
\usepackage{color}
\usepackage[colorlinks,linkcolor=blue,anchorcolor=blue,citecolor=blue]{hyperref}
\usepackage{cleveref}   
\usepackage{natbib}
\usepackage[T1]{fontenc}
\usepackage{wasysym}
\usepackage{times}
\usepackage{appendix}

\begin{document}
	
	
	\title{	The primordial black holes solution to the cosmological monopole problem}
	\author{Xin-Zhe Wang}
	\affiliation{Guangxi Key Laboratory for Relativistic Astrophysics, Department of Physics, Guangxi University, Nanning 530004, China}
	\author{Can-Min Deng}
	\email{dengcm@gxu.edu.cn}
	\affiliation{Guangxi Key Laboratory for Relativistic Astrophysics, Department of Physics, Guangxi University, Nanning 530004, China}

	\begin{abstract}
		Recently, the pulsar timing array (PTA) collaborations, including CPTA, EPTA, NANOGrav,  and PPTA, announced that they detected a stochastic gravitational wave background spectrum in the nHz band. This may be relevant to the cosmological phase transition suggested by some models. Magnetic monopoles and primordial black holes (PBHs), two unsolved mysteries in the universe, may also have their production related to the cosmological phase transition.
		Inspired by that, we revisit the model proposed by Stojkovic and Freese, which involves PBHs accretion to solve the cosmological magnetic monopole problem. We further develop it by considering the increase in the mass of the PBHs during accretion and taking the effect of Hawking radiation into account. With these new considerations, we find that solutions to the problem still exist within a certain parameter space. In {addition}, we also generalize the analysis to PBHs with {an} extended distribution in mass. This may be a more interesting scenario because PBHs that have accreted magnetic monopoles might produce observable electromagnetic signals if they are massive enough to survive in the late universe.
	\end{abstract}
	
	
	\maketitle
	
	\section{Introduction}
	{Since the advent of Maxwell’s theory of electromagnetism, magnetic monopoles have been the object of great interest. This is because, if magnetic monopoles are present, then Maxwell’s equations would have perfect electro-magnetic symmetry. Dirac came up with a better motivation, showing that the existence of a single magnetic monopole was sufficient to explain the quantization of the electric charge, i.e., all electric charges would be an integer multiple of the fundamental unit \citep{1931RSPSA.133...60D}. The expression derived by Dirac is called the Dirac quantization condition, which gave $ eg=n/2 $,  where $e$ is the electric charge, $g$ is magnetic charge of the monopoles, and n is an integer. After the advent of the grand unified theory (GUT), \cite{HOOFT1974276} showed that magnetic monopoles appeared as solutions to the field equations of the theory.  In this theory, the weak interaction and the strong interaction adopt a uniform behavior in the high energy range, that is, in a single symmetry group SU(5) \citep{georgi1974hierarchy}.  The generation of magnetic monopoles occurs in this unified spontaneous symmetry breaking {process. For example}, a simple gauge group SU(5) can be backed up as SU(3)$\otimes$SU(2)$ \otimes $U(1),  where  the U(1) factor would contain a point topology with magnetic charges \citep{PhysRevLett.43.1365}.  And this can be linked to the idea that in the early universe, magnetic monopoles may have formed at the time of GUT phase transition when energy distribution was very dense \citep{Preskill:1984gd}.
	The monopole mass satisfies $ M_m\geqslant4\pi m_v/e^2 $ where $ e=1/\sqrt{137} $ in cgs units and $ m_v $ is the intermediate vector boson mass \citep{vilenkin1994cosmic}. For a $ m_v\sim10^{14} $ GeV, one can estimate the monopole mass $ M_m\sim10^{16} $ GeV,  which is adopted in this work for calculation. According to the limitation of causality, one monopole per event horizon volume can be predicted. However, this would have led to the existence of excessive magnetic monopoles in the early universe \citep{kibble1976topology,PhysRevD.14.870,VILENKIN1985263,vilenkin1994cosmic,Stojkovic:2005zh}.
	Despite considering the monopole-antimonopole annihilation, the number density of residual magnetic monopoles is still unacceptable \citep{PhysRevLett.43.1365,ZELDOVICH1978239}. These theoretical considerations on magnetic monopoles have historically inspired a great deal of experimental {research} to find such particles. However, current astronomical observations limit the density of magnetic monopoles to a very low order of magnitude \citep{Freese:1983hz}. This contradiction implies the existence of a mechanism that {has} significantly reduced the density of magnetic monopoles, if a large number of magnetic monopoles {were} produced during the GUT phase transition in the early universe, which {is} known as the cosmological monopole problem.
    } 
    
	{Recently,  a new batch of pulsar timing (PTA) data from CPTA, EPTA, NANOGrav, and PPTA show an excess of a stochastic gravitational wave background spectrum in the nHz band \citep{NANOGrav:2023gor,Reardon:2023gzh,EPTA:2023fyk,Xu:2023wog}. In addition to the inspiraling of supermassive black hole binaries \citep{Burke-Spolaor:2018bvk,Sesana:2008mz}, there are many other cosmological processes that can serve as potential sources of gravitational waves in that observation frequency range. For example, it can be interpreted as the result of the  cosmological phase transition in some models \citep{Addazi:2023jvg,Nakai:2020oit,Wang:2023len,Fujikura:2023lkn,Megias:2023kiy,Bai:2023cqj,li2023primordial}. In addition, some authors suggested that such background gravitational waves may also have originated from the production of primordial black holes (PBHs) \citep{guo2023footprints,kitajima2023nanohertz,Franciolini:2023pbf}. 
	The PBHs were {thought} to be formed in the regions with high density which undergo gravitational collapse in the early universe due to primordial perturbations \citep{Zeldovich:1967lct,1971MNRAS.152...75H,2016PhRvD..94h3504C}, which may also occur during the cosmological phase transitions \citep{Liu:2021svg,Liu:2022lvz}. Detection of these gravitational wave signals may provide valuable clues for studying the physics of the early universe.
	}
	
	{Magnetic monopoles and primordial black holes are two mysteries in the universe. Their production may both be related to the cosmological phase transition. It is interesting to consider what would have happened if they had met in the early universe.} Stojkovic and Freese proposed that the PBHs with mass less than $ 10^9 $ g can sufficiently  capture the magnetic monopoles in the early universe, which may be  an alternative solution to the cosmological monopole problem \citep{STOJKOVIC2005251}. 
	In Ref.\cite{STOJKOVIC2005251}, the monochromatic mass for PBHs was  considered. However, it is likely that the mass function of PBHs could be extended \citep{2016PhRvD..94h3504C,2020ARNPS..70..355C}. If an extended mass function is {considered}, then those PBHs that are massive enough would not evaporate quickly and survive for a long time in the universe. Then, a very tempting question arises, do those PBHs that have captured a certain amount of {magnetic monopoles produce} any observable signals? If there are and people can detect these signals, it probably means that one has the opportunity to catch the shadow of the two mysteries at the same time. Interestingly, our previous study has found that those PBHs {holding} a certain amount of magnetic charges would strongly radiate electromagnetic waves during the process of merging with each other \citep{2018PhRvD..98l3016D}. In particular, the characteristics of the radio pulses produced by the merger of those PBHs with masses around the mass of the Earth are consistent with the recently discovered radio transient source i.e. nonrepeating fast radio bursts \citep{2021PhRvD.103l3030D}. More generally, it can be imagined that these electromagnetic signals radiated by the merger of those PBHs with different mass should also be different. Before one can figure that out, one needs to know how many magnetic charges the {PBHs with} various masses have {held} after the accretion. Therefore, it is of great interest to develop the idea proposed in Ref \cite{STOJKOVIC2005251}, and further derive the distribution of magnetic charges in the PBHs after the accretion. In Ref \cite{STOJKOVIC2005251}, the calculation is based on the assumption that the black hole's mass is constant during the accretion and does not take into account the Hawking radiation.
	In this paper, we develop this process by taking into account the increase in the mass of the PBHs during accretion, and consider the effect of Hawking radiation of the PBHs itself on the accretion, which makes the calculation more self-consistent. Furthermore, based on the results of this paper, {we} will investigate the electromagnetic radiation and its observability that may be produced by these magnetized PBHs as they orbit each other until they merge in an accompanying paper. In order to facilitate the application of the results of this paper in astronomical research, we adopt the  formula system commonly used in astronomy.
	

	\section{The PBHs accretion model  \label{Sec.2}}
	In this section, we revisit the PBH accretion model for the cosmological monopole problem proposed by \cite{STOJKOVIC2005251}, and develop it by further considering the increase in the mass of the PBHs during accretion and taking the effect of Hawking radiation into account. The constraints of observations on the model parameter space are also obtained. {Let's} start from the beginning.
	
	If only annihilation is considered as the way to eliminate the magnetic monopoles, the dynamic equation for the number density of the cosmological magnetic monopoles $n_{\rm{m}}$ is given by \cite{PhysRevLett.43.1365}
	\begin{equation}\label{eq: only annihilation}
		\frac{dn_m(t)}{dt}=- \kappa n_m^2(t)-3\frac{\dot{a}}{a}n_m(t)~.
	\end{equation}
	The first term on the right hand side represents the annihilation rate, where $ \kappa $ characterizes the annihilation efficiency. The second term is the impact of the adiabatic expansion of the universe, where $ a $ is the scale factor. 
	However, \cite{PhysRevLett.43.1365} pointed out that this would result in an unacceptable number density of residual magnetic monopoles in the current universe. 
	\cite{STOJKOVIC2005251} suggested that the one may consider the accretion of the PBHs on the magnetic monopole, so that the dynamic equation becomes
	\begin{equation}\label{eq: n_m}
		\frac{dn_{\rm{m}}(t)}{dt}=- \kappa n_{\rm{m}}^2(t)- \sigma n_{\rm{BH}} v_{\rm{m}}  n_{\rm{m}}(t) -3\frac{\dot{a}}{a}n_{\rm{m}}(t)~.
	\end{equation}
	The second term on the right hand side takes into account the black hole's capture of monopoles, and the monopole flux is given by $n_{\rm{m}}v_{\rm{m}}$, where $v_{\rm{m}}$ is the velocity. $n_{\rm{BH}}$ is the number density of the PBHs, and $ \sigma$ is the cross section for capture of the monopole by a black hole. {It turns} out that the PBHs can effectively capture the magnetic monopoles, and then the cosmological monopole problem {is} solved \citep{STOJKOVIC2005251}. It should be noted that the growth of black hole mass and the effects of Hawking radiation were ignored in their calculations. Obviously, it is reasonable to assume that the mass of the black hole remains constant during the accretion process when the total mass of PBHs is much larger than the total mass of the magnetic monopoles initially, that is, $\beta_0 \gg  1$ where $\beta (t)=\rho_{\rm{BH}} / \rho_{\rm{m}}$ and  $\beta_0= \beta (t_0)$. However, this solution means that a large amount of PBHs {are} required.

	In this paper, we investigate the case of $\beta_0 \ll 1$. We assume that $ \sigma= 4 \pi G^2M^2/v_{\rm{m}}^4$, {the cross section for spherically symmetric accretion, which scales as the square of the
	mass of the black hole and is widely adopted in the {literature} \citep{Zeldovich:1967lct,1976MNRAS.177...51L,1981A&A....94..306B,Custodio:1998pv,PhysRevD.66.063505,PhysRevD.71.104009,PhysRevD.71.104010}} 
    When $\beta_0 \ll  1$, the mass {of} PBHs will increase significantly due to the accretion of magnetic monopoles, and the rate of the PBHs with a mass M increase is given by
	\begin{equation}\label{eq: Monochromatic mass function accretion rate}
		\frac{dM}{dt}= \sigma  n_{\rm{m}}  M_{\rm{m}} v_{\rm{m}} , 
	\end{equation}
	here $ v_m \sim \sqrt{8kT/(\pi M_{\rm{m}})} $ is the magnetic monopole thermal velocity. 
	For Hawking radiation effect, it can be {ignored} when the accretion power is greater than the radiated power.
		\begin{figure}[htbp]
			\centering
			\subfloat{
				\includegraphics[width=0.8\linewidth]{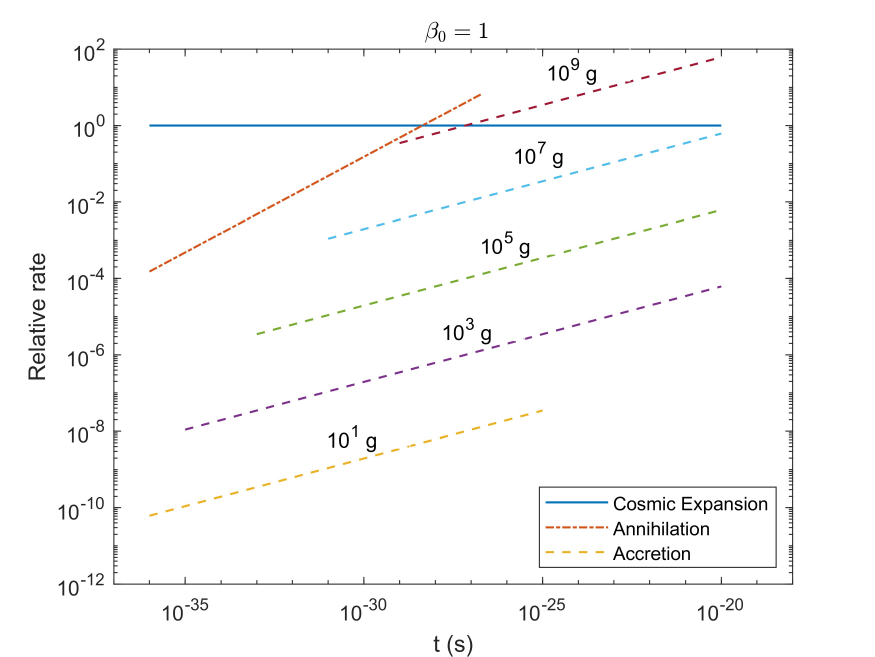}
			}\hfill
			\subfloat{
				\includegraphics[width=0.8\linewidth]{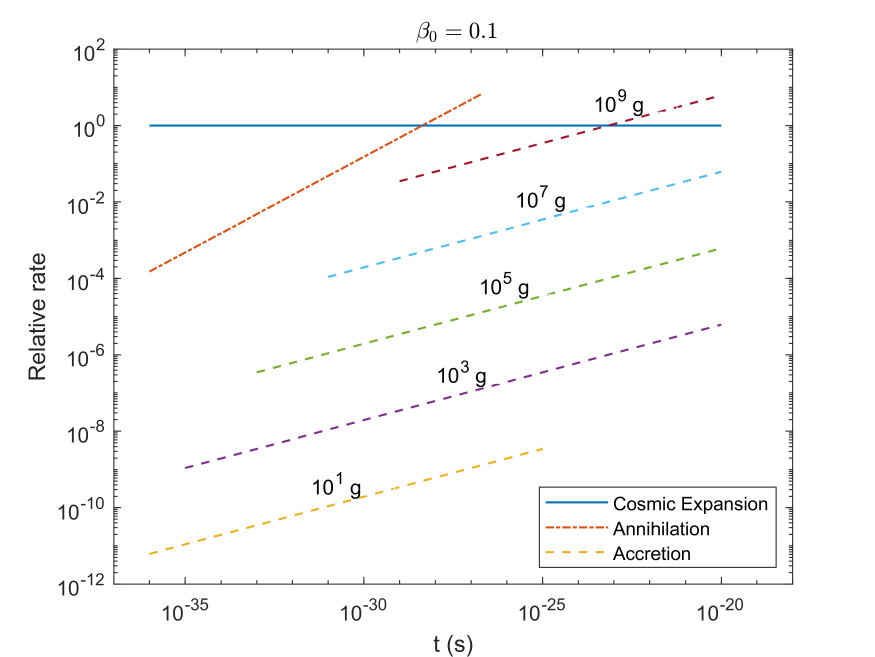}
			}\\
			\captionsetup{font=footnotesize}
		    \caption{The figures show the evolution {over time} of the contribution of cosmic expansion, annihilation, and accretion, with the cosmic expansion as a benchmark, to reducing the number density of the magnetic monopoles on the right side of Eq.(\ref{eq: n_m}).  Here the initial number density of the magnetic monopoles at the time of GUT phase transition  $\sim 10^{76}$ cm$^{-3}$ is adopted \citep{ryden2017introduction}.
		    As an example, the upper panel corresponds to the case of $\beta_0=1$, and the bottom panel corresponds to the case of $\beta_0=0.1$.}
			\label{figure:1}
		\end{figure}

		It can be seen that since the accretion increases the mass of the black hole and accelerates the accretion itself , the Eq.(\ref{eq: n_m}) is a nonlinear differential equation with no analytical solution. Therefore, we try to find out if there is an approximate solution. The annihilating term is crucial. As it is known, as the temperature $T$ drops, the annihilation of the magnetic monopoles will end at $ T_{\rm{c}} \simeq 10^{23} $ K \citep{PhysRevLett.43.1365}, which corresponds to the age of the universe $ t_{\rm{c}} \simeq 2.4 \times 10^{-27} $ s. As shown in figure \ref{figure:1}, the effect of annihilation on the reduction of magnetic monopole density is always greater than that of the accretion until annihilation ends, within the mass of the PBHs of interest. This means that one can divide the Eq.(\ref{eq: n_m}) into two parts at $t=t_{\rm{c}}$, where the accretion term is negligible  when $t<t_{\rm{c}}$, and the annihilation term disappears when $t>t_{\rm{c}}$. The situation of $t<t_{\rm{c}}$ {has} already been resolved by \cite{PhysRevLett.43.1365} analytically.
		In this work, we address those moments of $t>t_{\rm{c}}$.

		During the radiation universe, the Hubble's volume at time t evolves as $ V_{\rm{H}}  \varpropto t^{3/2} $, which gives the PBHs' number density $ n_{\rm{BH}}(t)\varpropto N_{\rm{BH}} t^{-3/2} $ {before they evaporate}, where $ N_{\rm{BH}} $ is the number of PBHs. Since the annihilation term disappears  for $t>t_{\rm{c}}$, and the cosmic expansion would not reduce the magnetic monopoles,  the rate of decline of the magnetic monopoles is equal to the accretion rate by the PBHs
		\begin{equation}\label{eq: MMF Number equation of magnetic monopoles}
			\frac{dN_{\rm{m}}(t)}{dt}= -  \sigma  n_{\rm{m}}  v_{\rm{m}}  N_{\rm{BH}},
		\end{equation}
		where $ N_{\rm{m}} =n_{\rm{m}} V_{\rm{H}}  $ is the number of monopoles.
		Dividing  E.q. (\ref{eq: MMF Number equation of magnetic monopoles}) by E.q. (\ref{eq: Monochromatic mass function accretion rate}), one gets
		\begin{equation}\label{eq: MMF Number equation of magnetic monopoles2}
			\frac{dN_m}{dM}=-\frac{N_{bh}}{M_m}~.
		\end{equation}
		Then, the solution of the magnetic monopole density can be obtained
		\begin{equation}
			{\label{eq: MMF magnetic monopoles number density and PBH mass}}
			n_m(t)=(1+\beta_{\rm{c}}-\frac{M}{M_{\rm{i}}} \beta_{\rm{c}} )~n_{\rm{c}} \left(\frac{t}{t_{\rm{c}}}\right)^{-3/2}~,
		\end{equation}
		where $n_{\rm{c}}= n_m(t_{\rm{c}})$ is number density of the  magnetic monopoles right at the end of the annihilation process, and  $\beta_{\rm{c}}= \beta(t_{\rm{c}})$ is the ratio of the total mass of PBHs to the total mass of the remained magnetic monopoles at $t_{\rm{c}}$ as an unknown parameter. We adopt the result of $n_{\rm{c}} \simeq 2.3 \times 10^{62} $ cm$^{-3}$ given by  \cite{PhysRevLett.43.1365}. The evolution of the mass $M(t)$ of the black hole is determined by 
		\begin{equation}
			{\label{eq: MMF Solutions of differential equations}}
			\begin{aligned}
				&\frac{M_{\rm{i}}}{M}-1+\frac{\beta_{\rm{c}}}{1+\beta_{\rm{c}}}\ln\left[\left(1+\beta_{\rm{c}}\right)\frac{M_{\rm{i}}}{M}-\beta_{\rm{c}}\right]\\
				&=\frac{4 \dot{M_{\rm{c}}} t_{\rm{c}}}{M_{\rm{i}}} (1+\beta_{\rm{c}})[1-\left(\frac{t}{t_{\rm{c}}}\right)^{1/4}]~,
			\end{aligned}
		\end{equation}
		where $\dot{M_{\rm{c}}}=\dot{M}(t_{\rm{c}})$ is given by E.q. (\ref{eq: Monochromatic mass function accretion rate}).
		As one can see, this solution cannot be reduced to an explicit function about $t$, but it can be written as an explicit function about $M$. As mentioned earlier, mass loss of the black holes caused by Hawking radiation should be considered, when the accretion rate decreases to close to the  power of Hawking radiation as the magnetic monopole density decreases. We define $t_{\rm{f}}$ as the time when the accretion power is equal to the power of the Hawking radiation $P_{\rm{H}}={\alpha \hbar c^6}/( G^2 M^2)$, at which time the mass of the black hole is denoted as $M_{\rm{f}}$, where $\alpha=4.3 \times 10^{-4}$ is adopted for $M<10^{14}$ g \citep{PhysRevD.13.198}. The $t_{\rm{f}}$ and $M_{\rm{f}}$ can be obtained numerically by equating E.q.(\ref{eq: Monochromatic mass function accretion rate}) with the power of the Hawking radiation, and combining E.q.(\ref{eq: MMF magnetic monopoles number density and PBH mass}) and E.q.(\ref{eq: MMF Solutions of differential equations}). Once  $t_{\rm{f}}$ and $M_{\rm{f}}$ are obtained, one can bring them into E.q.(\ref{eq: MMF magnetic monopoles number density and PBH mass}) to get $n_m(t_{\rm{f}}) $.
		It can be seen that the solution given by E.q.(\ref{eq: MMF magnetic monopoles number density and PBH mass}) and E.q.(\ref{eq: MMF Solutions of differential equations}) is valid for $t<t_{\rm{f}}$, and when $t>t_{\rm{f}}$ the number density evolves with the cosmic expansion $n_m(t) = n_m(t_{\rm{f}}) (t/t_{\rm{f}})^{-3/2}$ because the accretion can be ignored \footnote{When the Hawking radiation power is greater than the accretion power, the accretion would be  suppressed because the monopoles would be pushed away by the radiation pressure.}.

		Since E.q.(\ref{eq: MMF magnetic monopoles number density and PBH mass}) gives an implicit solution, it is useful to find its asymptotic form. 
		Indeed, for $ \tau \equiv t/t_{\rm{c}} \gg 1$,  E.q.(\ref{eq: MMF magnetic monopoles number density and PBH mass}) can be reduced to
		
		\begin{equation}
			{\label{eq: 2}}
			n_m(\tau) \simeq n_{\rm{c}}(1+1/\beta_{\rm{c}})~ e^{({1/\beta_{\rm{c}}-A \tau^{1/4}})}  \tau^{-3/2}~,
		\end{equation}
		where $ A={4  (1+\beta_{\rm{c}})^2  \dot{M_{\rm{c}}} t_{\rm{c}}}/(\beta_{\rm{c}} M_i) $. We are not going to discuss this solution in detail for  $\beta_{\rm{c}} \gg 1$, because it means that much more PBHs are needed. And it's not unusual that there are enough PBHs to solve the magnetic monopole problem. We are more interested in the case of $\beta_{\rm{c}} \ll 1$. 
		In this case, is there still a solution that meets the observation limits? That is, in order to avoid a conflict with observations whether there is a set of parameters $(M_i, \beta_{\rm{c}})$ such that $n_m(t_{\rm{f}}) = n_m(t_{\rm{age}}) (t_{\rm{age}}/t_{\rm{f}})^{3/2}$, where the upper limit on monopole number density in the current universe  is $n_m(t_{\rm{age}}) < 10^{-25} \rm{cm}^{-3}$ \citep{PhysRevLett.51.1625},  and an age $t_{\rm{age}}=$ 13.8 billion years is adopted.  After some calculations, we found that such a solution does exist.  For example, $M_i=10^{6}$ g and $\beta_{\rm{c}}=10^{-3}$, which results in $\tau_{\rm{f}} \equiv  t_{\rm{f}}/ t_{\rm{c}}=3 \times 10^8$, and bringing $t_{\rm{f}}$ into E.q.(\ref{eq: 2}) shows that this solution satisfies the observation limit. 
		It can be seen that a relatively small number of PBHs can also solve the magnetic monopole problem, and we call such solutions as non-trivial solutions.

\begin{figure}[htbp]
			\centering
			\includegraphics[width=0.8\linewidth]{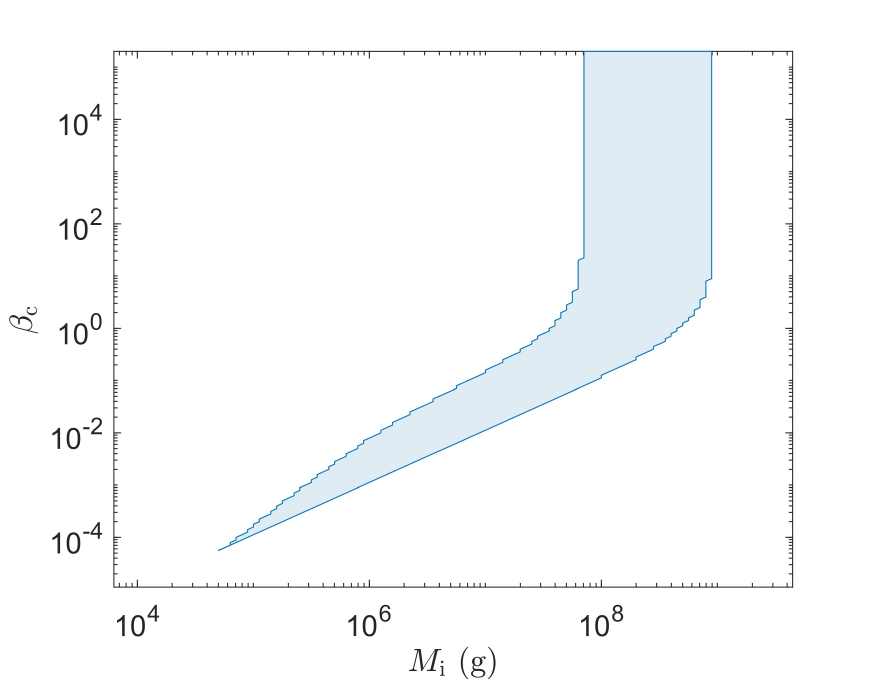}
			\captionsetup{font={scriptsize}}
			\caption{The blue area is the parameters space of the solution which meets the  astronomical observation constrain for $t_{\rm{b}} \leqslant t_{\rm{c}}$. The upper bound corresponds to the consequence of requiring that the density of the PBHs cannot be greater than the  energy density of the universe when they are produced.}
			\label{figure: 2}
		\end{figure}

		In order to obtain the full parameter space that does not violate  the observation limits, it is necessary to solve E.q.(\ref{eq: MMF magnetic monopoles number density and PBH mass}) and E.q.(\ref{eq: MMF Solutions of differential equations}) directly. But before we do that, let's have some  analysis. First of all, the accretion power of the black hole is required to be  significantly greater than the Hawking radiation power at $t_{\rm{c}}$, otherwise the accretion will be greatly suppressed  at the beginning. For PBHs of $M_i > 10^{2}$ g, this requirement is met, in which case it is reasonable to ignore the Hawking radiation at the beginning, as in E.q.(\ref{eq: n_m}).
		Second, in addition to meeting the observation constrain of the magnetic monopoles, the black hole needs to evaporate before the primary nucleosynthesis, otherwise it will seriously affect the nucleosynthesis. This requires $M_i \lesssim 10^{9}$ g. Taking the above requirements into account, we finally get the full parameter space, as shown in figure  \ref{figure: 2}. One can see that there are solutions in the interval $10^5 {\rm{g}} \lesssim M_i \lesssim 10^8 {\rm{g}}$ for  $\beta_{\rm{c}}<1$  and $10^7 {\rm{g}} \lesssim M_i \lesssim 10^9 {\rm{g}}$ for  $\beta_{\rm{c}}>1$. 
		\par

		\section{Effect of the production time of the PBHs \label{Sec.3}} 
		
\begin{figure}[htbp]
			\centering
			\subfloat{
				\includegraphics[width=0.8\linewidth]{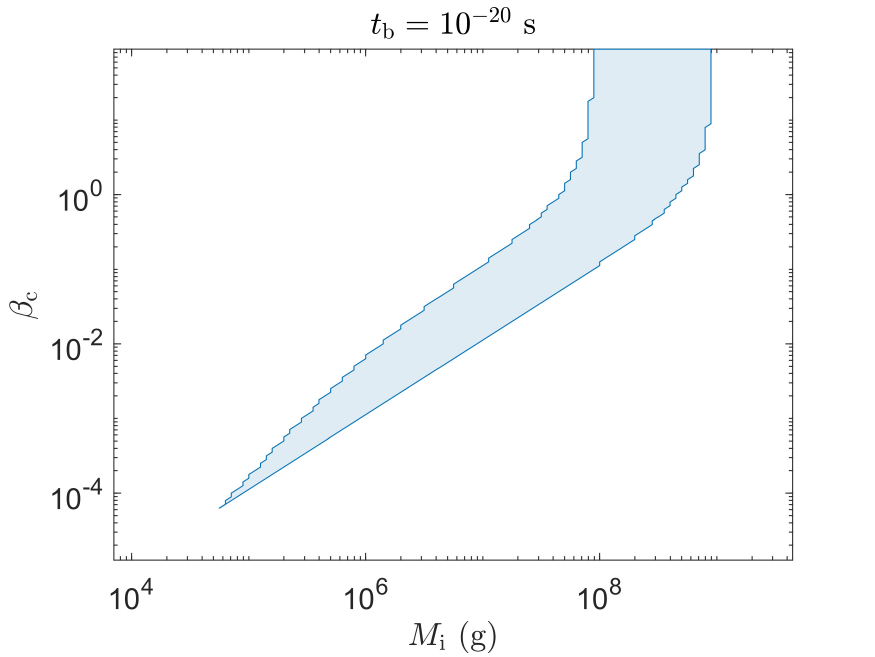}
			}\hfill
			\subfloat{
				\includegraphics[width=0.8\linewidth]{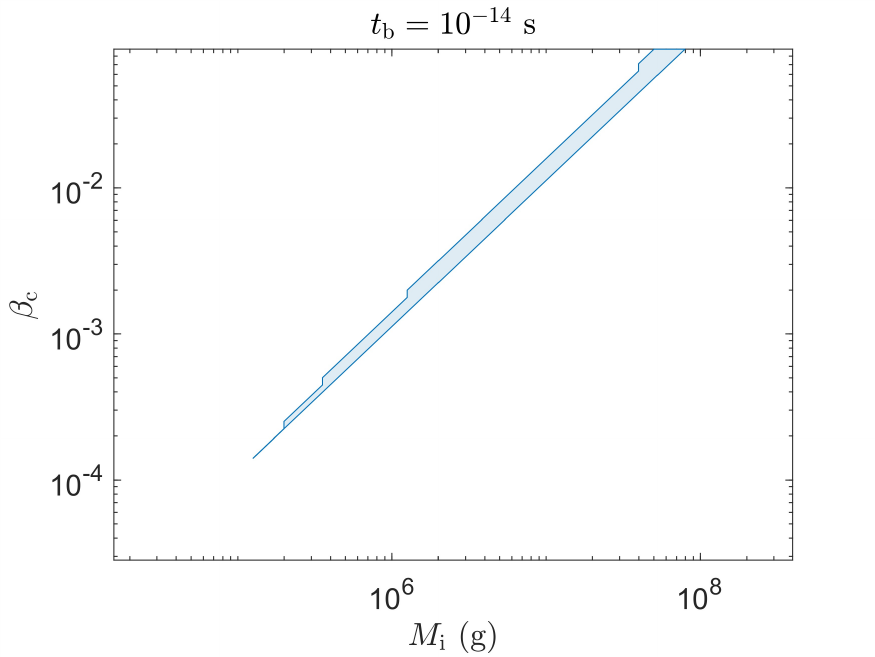}
			}\\
			\captionsetup{font=footnotesize}
			\caption{The meaning of the blue area is the same as that in figure \ref{figure: 2}, which only corresponds to different generation time of the PBHs. The upper and bottom panels correspond to $ t_{\rm{b}}=10^{-20}$ s and $ t_{\rm{b}}=10^{-14}$, respectively, as examples.}
			\label{figure:3}
		\end{figure}

		 It should be noted that there is an implicit assumption in the previous discussion, that is $t_{\rm{b}} \leqslant t_{\rm{c}}$, where  $t_{\rm{b}}$ the time when the PBHs were produced.  If it is the opposite  $t_{\rm{b}} > t_{\rm{c}}$, then the solution for  $n_m(t)$  will be different. Specifically, the magnetic monopoles undergo one more phase of cosmic expansion in the period between the end of annihilation and the beginning of the accretion ($t_{\rm{c}} < t <  t_{\rm{b}}$,) than was discussed in the previous section.  In this case, the parameter space of the solution to the monopole problem is shown in figure \ref{figure:3}.  As one can see, the parameter space of the solution  shrinks as $ t_{\rm{b}}$ increases, and there is no solution for $ t_{\rm{b}} \gtrsim 10^{-13}$ s.
		\par

		\section{ PBH with Extended mass function \label{Sec.4}}
        So far, all of our discussions have been based on the assumption that the PBHs are monochromatic in mass.
		Although the monochromatic mass assumption of PBHs has been adopted by many authors, there are many proposed mechanisms indicating that PBHs may have an extended mass function \citep{PhysRevD.60.063509,Poulter:2019ooo,PhysRevLett.121.081306}. Since the PBHs are considered as a candidate for DM  \citep{PhysRevD.105.103508,Bellomo_2018,PhysRevD.94.083504,Green_2021,PhysRevD.96.123523,PhysRevD.87.123524,PhysRevD.96.043504,PhysRevLett.111.181302,PhysRevD.101.123514,10.1093/mnras/stac2705}, we denote $f$ as the mass fraction of the PBHs to the dark matter in the current universe.
		We adopted the power law form for the mass function of the PBHs considered by \citep{PhysRevD.96.023514}
		\begin{equation}
			f \varpropto M^{\gamma-1},   \quad M_\text{min}  \leqslant M\leqslant M_\text{max},
		\end{equation}
		where $ \left\{ \gamma||{\gamma}|<1,\gamma\neq0 \right\} $, $M_\text{max}$ may correspond to the horizon mass at the collapse epoch, $M_\text{min}$ is the cuttoff mass in the low mass end.

		For simplicity, we adopt a discrete distribution $f_i \varpropto M_i^{\gamma-1}$ as an approximation, 
		where $ f_i$ is the initial mass fraction of the PBH with a mass $M_i=10^i$ g to the dark matter. 
		In this regard, the dynamic equation can be written as $(t \geqslant t_{\rm{c}})$
		\begin{equation}\label{eq: 10}
			\frac{dn_m(t)}{dt}=-\sum\limits_{i} X_i-3\frac{\dot{a}}{a}n_{\rm{m}}(t)~.
		\end{equation}
		where $\sigma_i=4\pi G^2{M_i}^2/{v^4_m}$ and $X_i= \sigma_i n_{{\rm{BH}},i} v_m n_m(t)$.
		Apparently, $X_i$ represents the contribution of PBHs  with mass of $M_i$. 
		As one can imagine, the dynamic equation in the case of  extended mass distribution is much more complicated than that of the monochromatic case. Therefore, we need to do the following analysis to figure out the problem. Let's consider it this way, looking at the following ratios,
		\begin{equation}\label{eq: X_i/X_j}
			\frac{X_i}{X_j}=\left(\frac{M^0_{i}}{M^0_{j}}\right)^{\gamma}\left(\frac{M^0_{j}/M_j}{M^0_{i}/M_{i}}\right)^2~,
		\end{equation}
		here, we take $i$,$j$ satisfy $M^0_{i}<M^0_{j}$, and the superscript of $0$ indicates the  initial value of the  physical quantity.  Let's first look at the initial time, where ${X_i}/{X_j}=({M^0_{i}}/{M^0_{j}})^{\gamma}$,  and obviously one can find that ${X_i}/{X_j}>1$ for $\gamma<0$, and ${X_i}/{X_j}<1$ for $\gamma>0$.
		On the other hand, the mass of the black holes is increasing due to accretion, so that $M_{i}>M^0_{i}$.
		In the case of $\gamma<0$, since PBHs with a smaller mass $M_i$ dominate the accretion at the initial moment, it will also dominate the accretion at any subsequent time because ${X_i}/{X_j}>1$ always holds due to accretion such that $M^0_{j}/M_j > M^0_{i}/M_{i}$.
		Therefore, for  $\gamma<0$, the accretion is always dominated by the PBHs that $M_i=M_{\rm{min}}$. 
		In the case of $\gamma>0$, the situation is more complicated, and the conclusion cannot be judged directly from Eq.(\ref{eq: X_i/X_j}). However, when one derives the Eq.(\ref{eq: X_i/X_j}), one finds that its derivative $d({X_i}/{X_j})/dt$ is always $< 0$ (see Appendix \ref{appendix: A}). So one can conclude that since PBH with a larger mass $M_j$ dominates accretion at the initial moment for $\gamma>0$, it will also dominate accretion at any subsequent time because ${X_i}/{X_j}<1$ always holds during the  accretion.
	   Therefore, for  $\gamma>0$, the accretion is always dominated by the PBHs that $M_j=M_{\rm{max}}$. 
		\par
		
\begin{figure}[htbp]
			\centering
			\subfloat{
				\includegraphics[width=0.8\linewidth]{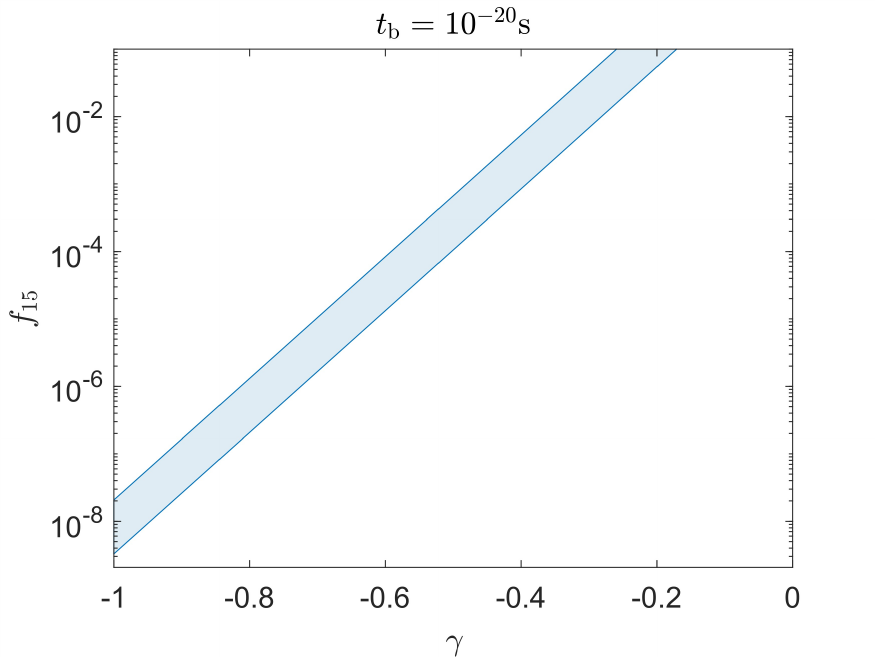}
 			}\hfill
			\subfloat{
				\includegraphics[width=0.8\linewidth]{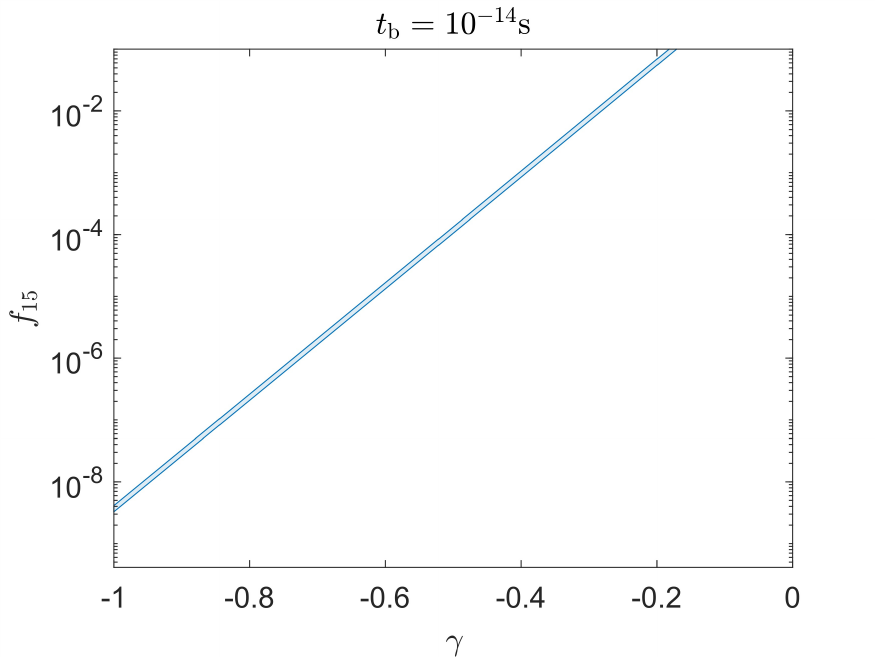}
			}\\
			\captionsetup{font=footnotesize}
				\caption{The blue area represents the parameter space that meets the astronomical observation {constraints} for a given $t_{\rm{b}}$. Since the accretion in the power-law case is dominated by either the PBHs with the largest mass or the PBHs with the smallest mass, the parameter space  appears  as a narrow band. As one can see, {for a larger} $t_{\rm{b}}$, the narrower the parameter space as in the monochromatic case (see figure \ref{figure:3}).
				}
			\label{figure: 4}
		\end{figure}

		Therefore, we conclude that the population of PBHs that dominates the accretion did not change over time, and one can apply the results of the previous section on monochromatic PBHs to the problem here.
		In this way, one can find the solutions of  Eq.(\ref{eq: 10}). The solutions are $(M_{\rm{min}},\beta_0)\in\varXi$ for $\gamma<0$, and $(M_{\rm{max}},\beta_0)\in \varXi$ for $\gamma>0$, where $\varXi$ is the parameter space of the solution for the monochromatic PBHs as shown in figure \ref{figure: 2} $(t_{\rm{b}} = t_{\rm{c}})$ for example. 	In the case of $\gamma>0$, there would be no PBHs that survive in the current universe under the solutions of $(M_{\rm{max}},\beta_0)\in \varXi$. 
		However, in the case of $\gamma<0$, because the solutions of $(M_{\rm{min}},\beta_0)\in\varXi$ has no restrictions on $M_{\rm{max}}$, there might be PBHs with masses $\gtrsim 10^{15}$ g surviving to this day as candidates for dark matter.
		Therefore, we are concerned with the total mass of these surviving PBHs, and require that $f_{15}=\sum\limits_{i \geqslant 15} f_i$ must be $<1$. The parameter space of $(\gamma,f_{15})$ allowed by this solution is shown in figure \ref{figure: 4}.
		
		\section{Summary and discussion \label{Sec.5}}	
		In this paper, we develop the PBHs accretion model to the cosmological monopole problem for the case of $\beta_0 \ll 1$, and  derive the analytical solution.  In {addition}, the parameters space of the solution, which meets  the  astronomical observation {constraints} for the magnetic monopoles, is also {obtained}. We further generalize this model successfully to the cases where the PBHs' mass {has} an extended distribution. It should be noted that the spherical accretion is {considered} in this paper instead of the simple gravitational capture process used in \cite{STOJKOVIC2005251}, resulting in a stronger ability of the PBHs to capture the magnetic monopoles. If we adopt the gravitational capture modelling as in \cite{STOJKOVIC2005251}, we find that there is still a solution in our calculations, although the parameter space for the solution is significantly reduced.

		It is worth pointing out that our purposes in doing this work is not only to solve the {cosmological} magnetic monopole problem itself.
		In the context of the model, it would be possible for PBHs, which would naturally carry a certain amount of magnetic monopoles, with large enough mass to survive into the late universe: Firstly, even if the PBHs were monochromatic in mass, these PBHs could easily merge with each other frequently in the early universe due to the high number density, and the mass of the black holes after multiple mergers may be large enough to survive until the late universe; secondly, if the PBHs have an extended mass distribution, in addition to the merger effect to produce the massive black holes, there are massive  PBHs in the large end of the mass function. When these PBHs carrying magnetic monopoles move in the universe, according to electrodynamics, they may produce strong electromagnetic radiation under suitable conditions, such as orbiting. In our previous works, we have proposed the possibility and observability of the electromagnetic radiation produced by these PBHs when they merge with each other \citep{2018PhRvD..98l3016D,2021PhRvD.103l3030D}.
		In the future, if people can actually detect the electromagnetic signals produced by those PBHs, it may be of great physical and astronomical interest. For a detailed discussion in this regard, we're going to present it in the next paper based on the results of this work. The magnetic monopoles and primordial black holes are both living fossils of the Big Bang, and we believe that by studying their interaction and the subsequent possible observational effects, we are hopeful to find a new probe into the physics of the early universe.

		\medskip\noindent\textit{Note added\,---\,}
		As this work was nearing completion, we found new papers discussing similar topics  on Arxiv \citep{Zhang:2023tfv,Zhang:2023zmb}. Their calculations differ from ours by still assuming that the PBH's mass remains constant during the whole process, as well as taking a {diffusive} capture model which makes it less efficient for PBHs to capture magnetic monopoles, and also mandating that the density of the matter cannot exceed the radiation in the early universe which {may not be necessary}.  Based on those three assumptions,  they concluded that  the monopole problem cannot be solved by PBHs capture. In our point of view, those three assumptions had a significant impact on the results. For example, the capture of magnetic monopoles by PBHs results in an increase in the mass, and this increased mass, in turn, enhances the PBH's capturing capability. This creates a positive feedback loop, leading to an exponential growth in the PBHs' ability to capture the magnetic monopoles. Therefore, the increase in the PBHs' mass is an effect that cannot be ignored.

    	\section*{Acknowledgments}
		We thank Dejan Stojkovic for useful discussions. 
		This work is supported by the National Natural Science Foundation of China (grant No. 12203013), and the Guangxi Science Foundation (grant Nos.2023GXNSFBA026030  and AD22035171). $\\$

		\section*{appendix}
			\begin{appendices}
			\setcounter{table}{0}
			\setcounter{figure}{0}
			\setcounter{equation}{0}
			\renewcommand{\thetable}{\thesection-\arabic{table}}
			\renewcommand{\theequation}{\thesection-\arabic{equation}}
			\renewcommand{\thefigure}{\thesection-\arabic{figure}}
			\section{PBH with the maximum mass dominates the accretion\label{appendix: A}}
			The accretion rate of magnetic monopoles  is given by Eq.(\ref{eq: Monochromatic mass function accretion rate}),  and integrating it, one gets
			\begin{subnumcases}{\label{eq: EMF PBH Accretion rate}}
				M_i=\left[{M_{i0}}^{-1}-\varphi(t)\right]^{-1}\\
				\varphi(t)=\int_{t_0}^{t}\frac{4\pi G^2\rho_m(t)}{v^3(t)}~.
			\end{subnumcases}
			Combined with  Eq.(\ref{eq: X_i/X_j}), one obtains
			\begin{equation}
				\frac{X_i}{X_j}=\left(\frac{M_{i0}}{M_{j0}}\right)^{\gamma-2}\frac{\left[{M_{i0}}^{-1}-\varphi(t)\right]^{-2}}{\left[{M_{j0}}^{-1}-\varphi(t)\right]^{-2}}~.
			\end{equation}
			At the initial time, one has $X_i/X_j<1$ for $\gamma>0$ and $X_i/X_j>1$ for $\gamma<0$.
			Taking the derivative of $X_i/X_j$ yields
			\begin{equation}
		\begin{aligned}
		\frac{d}{dt}\left(\frac{X_i}{X_j}\right)=
		& 2\alpha \varphi(t)\frac{\left[{M_{i0}}^{-1}-\varphi(t)\right]^{-2}}
		{\left[{M_{j0}}^{-1}-\varphi(t)\right]^{-2}} \\
	    &\left\{\left[{M_{i0}}^{-1}-\varphi(t)\right]^{-1}-\left[{M_{j0}}^{-1}-\varphi(t)\right]^{-1}\right\}<0~,
		\end{aligned}
		\end{equation}
		where $\alpha=(M_{i0}/M_{j0})^{\gamma-2}$. Therefore, the accretion would be always dominated by the PBHs with maximum mass in the case of $\gamma>0$.
		\end{appendices}

   	\bibliography{References}

	\end{document}